\documentclass{iucrjournals}

\title{
    Automating the analysis of micron-scale synchrotron diffraction data on inhomogeneous polycrystalline samples: 
    a solid oxide electrolysis cell case study
}

\author[a]{
    Liam A. V. Nagle-Cocco
    \footnote{Email: lnc@slac.stanford.edu}\IUCrCemaillink{lnc@slac.stanford.edu}
    \IUCrOrcidlink{0000-0001-9265-1588}
}
\author[a]{Christopher A. Crain\IUCrOrcidlink{0009-0002-9281-3254}}
\author[b]{Michael J. Dzara\IUCrOrcidlink{0000-0001-8125-0586}}
\author[a]{Mathias A. Kiefer}
\author[a,c]{Madeline G. Port\IUCrOrcidlink{0009-0003-1682-4338}}
\author[a]{Tolga Han Ulucan\IUCrOrcidlink{0000-0003-2701-4034}}
\author[b]{Madeline Van Winkle\IUCrOrcidlink{0000-0001-5772-2723}}
\author[b]{Oscar Hathaway}
\author[a]{
    Nicholas A. Strange
    \footnote{Email: nstrange@slac.stanford.edu}\IUCrCemaillink{nstrange@slac.stanford.edu}
\IUCrOrcidlink{0000-0001-5699-7274}
}

\affil[a]{Stanford Synchrotron Radiation Lightsource, SLAC National Accelerator Laboratory, Menlo Park, 94025, United States of America}
\affil[b]{Materials, Chemical, and Computational Sciences Directorate, National Laboratory of the Rockies, Golden, 80401, United States of America}
\affil[c]{Department of Materials Science and Engineering \& Department of Chemistry, University of Pennsylvania, Philadelphia, Pennsylvania 19104, United States of America}

\begin{document} 
\maketitle 

\begin{synopsis}
    An automated approach is presented for the analysis of micron-focused, spatially-resolved synchrotron diffraction data on the cross-section of chemically inhomogeneous polycrystalline samples such as batteries, solid oxide fuel/electrolysis cells, and photovoltaic devices. 
    A case study is shown, applying the approach to solid oxide electrolyser cells for a \textit{post mortem} characterisation. 
\end{synopsis}

\begin{abstract}
    A novel approach to \textit{post mortem} characterisation of electrochemical and photovoltaic devices is spatially-resolved diffraction using a hyper-focused, micron-width x-ray beam to examine the distribution of degradation products and strain, a technique called $\mu$-XRD. 
    Aside from the experimental difficulties associated with beam focusing and sample preparation, which are themselves non-trivial, the analysis of resulting data is complex and challenging, with full Rietveld analysis rarely attempted in literature. 
    The difficulty lies in the size of the data, which may consist of hundreds or even thousands of diffraction patterns with very different crystallographic phase compositions depending on position within the device, and the difficulty in fitting the data due to the presence of many phases at the same position, including possible degradation products which may be difficult to index and assign to known phases. 
    In this paper, we present a fully-automated open access Python routine for performing phase identification and Rietveld analysis on 2D datasets of diffraction pattern taken at micron-scale positions, measured over the cross-section of a chemically inhomogeneous device with polycrystalline phases.
    Solid oxide electrolyser cells (SOECs) are a promising technology for green hydrogen production which can utilise waste heat to split water at higher efficiencies than low-temperature electrolysis techniques such as polymer electrolyte membranes (PEM), but exhibit many degradation modes due to the high operating temperatures. 
    We present a case study using our analysis protocol on an SOEC fragment encompassing the air electrode, cation diffusion barrier, electrolyte, and fuel electrode. 
    With modification, this protocol could be applied to other devices such as all-solid-state batteries, wet-electrolyte battery electrodes, solid oxide fuel cells, photovoltaic devices, and metal-oxide pseudocapacitors. 
\end{abstract}


\clearpage
\section{Introduction}

\begin{figure}[h]
    \includegraphics[scale=0.99]{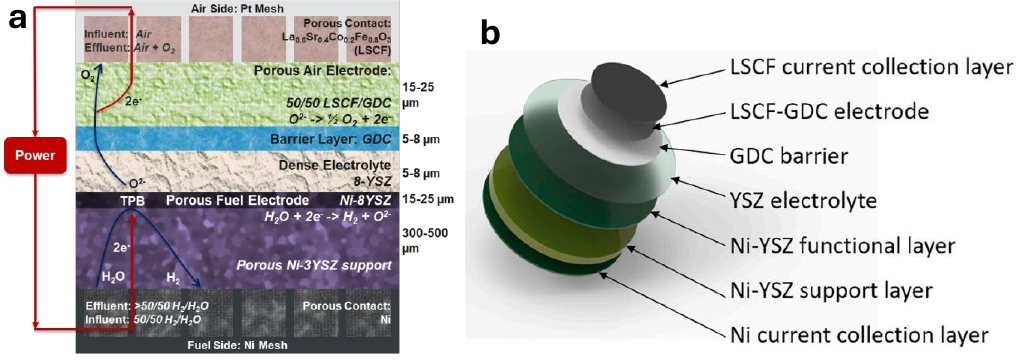}
    \caption{
        \label{SOEC_diagram.pdf}
    (a) Schematic illustrating the composition and design of the SOECs in this study. 
    (b) Schematic showing the construction of the coin cells used in this study, reproduced from \protect\citename{zhu2025voltage} \protect\citeyear{zhu2025voltage}. 
    }
\end{figure}

Many functional energy materials and devices are intrinsically chemically inhomogeneous, comprising multiple, spatially distinct regions with different compositions, crystal structures, and microstructures. In layered electrochemical and energy-conversion devices such as solid oxide cells, batteries, photovoltaic devices, and metal--oxide pseudocapacitors, there are different components (for example, electrodes, electrolytes, current collectors, and interlayers) which may contain largely mutually exclusive phase assemblages, while interfaces can host additional reaction products or structurally modified material. Superimposed on this architectural inhomogeneity are operating gradients in temperature, chemical potential, and electrochemical potential, which further drive spatially varying defect chemistry, strain, and secondary phase formation. Meaningful characterisation of such systems therefore requires techniques capable of resolving crystallographic information as a function of position, bridging the gap between bulk diffraction methods and highly local probes such as electron microscopy.

Powder X-ray diffraction (XRD) is generally considered a bulk technique, insensitive to local structural information. 
However, when an incident X-ray beam is focused, the illuminated volume is reduced and observed diffraction can be associated with a subsection of a sample. 
This capability is critical for SOECs where component layer thicknesses are on the order of micrometers. 
When X-rays are focused to the micron scale, there are typically efficiency losses on the order of 5\%--20\% depending on beamline optics~\cite{kim2018focusing,tang2019micro,yumoto2025x}. 
These losses can be compensated for by using a high flux undulator insertion device at a synchrotron X-ray facility, such as beamline 17-2~\cite{Hong2025RCI} at the Stanford Synchrotron Radiation Lightsource (SSRL), a technique called $\mu$XRD. 
Given these capabilities, rapid data acquisition is possible with a micro-focused beam, enabling two-dimensional mapping of XRD across materials at the micron-scale. 

A limiting factor in the ability to perform these experiments and maximise the information gain from measured data is in the automation of analysis. 
A measurement using a 1\,$\mu$m$^2$ beamsize, measuring at all positions in a 100\,$\mu$m$\times$100\,$\mu$m grid, will contain 10$^4$ diffraction patterns. 
Traditionally, diffraction data is analysed by Rietveld refinement~\cite{rietveld1969profile} which requires prior knowledge regarding the phases that are present; such refinements may be performed sequentially but this cannot account for chemical variations over the grid as the phases present will vary. 
A comprehensive treatment of a large dataset, including identification of the presence and extent of impurity phases, will require an unfeasible level of effort to manually analyse over all positions. 
Current approaches to this problem include the software package \textsc{XRDUA}~\cite{de2014xrdua}, which includes the option for Rietveld analysis, and more recently a deep learning algorithm with a multitask learning (MTL) architecture~\cite{li2024accurate}.

In this work, we present a different approach to fully automating the analysis of spatially resolved $\mu$XRD datasets measured over chemically inhomogeneous samples, using the well-established \textsc{Topas} software~\cite{Coelho2018TOPASAn} implemented via Python scripts~\cite{van1995python}. 
The protocol combines adaptive phase-identification strategies with automated Rietveld and Pawley refinement, enabling comprehensive analysis of large two-dimensional diffraction datasets based on user-determined candidate phase lists, without prior knowledge of local architecture. 

Solid oxide electrolyser cells (SOECs) provide a relevant case study for demonstrating this approach. Among green hydrogen production technologies, high-temperature electrolysis offers reduced electrical energy input relative to low-temperature electrolysis methods such as polymer electrolyte membranes~\cite{hauch2020recent}, but remains limited by material durability. 
Oxygen-conducting SOECs are composed of layered metal-oxide components that must withstand high temperatures, applied potentials, and large oxygen chemical potential gradients; a typical SOEC is indicated in Figure~\ref{SOEC_diagram.pdf}. 
These conditions give rise to a range of degradation mechanisms, including cation migration and secondary phase formation in air electrodes~\cite{lu2017srzro3,rowberg2024impact}, oxidation of Ni in fuel electrodes~\cite{neidhardt2019microkinetic}, and contamination from ancillary cell components~\cite{perz2016long,wei2015chromium,liu2024onset}. 

$\mu$XRD has been performed before~\cite{nozaki2020microstructure} on similar solid oxide cells, reporting subtle spatial variations in lattice parameters of both LSCF-6428 air electrode and GDC from Rietveld refinement, owing to the diffusion of atoms and changes in vacancy concentrations in the cell. 
However, this study involved Rietveld refinement only of several points along a linescan, missing out on the opportunity to gain information on variations within layers. 
To our knowledge, there have been no studies on SOECs which make use of spatially-resolved, micron-scale XRD in more than one dimension, or with comprehensive multi-phase Rietveld refinement over all positions. 

Here, we apply the automated analysis protocol presented in this paper to $\mu$XRD datasets measured across the cross-sections of operated and baseline oxygen-conducting SOEC fragments. 
While SOECs serve as a demanding validation case, with modification the protocol could be applied to a wide range of electrochemical and energy-conversion devices exhibiting a distribution of polycrystalline phases, and likely also in other fields from mineral studies~\cite{flemming2007micro} to cultural heritage~\cite{gonzalez2020x}. 
By removing a major barrier to comprehensive analysis of $\mu$XRD data, this work aims to facilitate broader adoption of spatially resolved diffraction as a routine tool for the study of chemically inhomogeneous energy conversion devices.

\section{Analysis protocol}

\begin{figure}[t]
    \includegraphics[width=\linewidth]{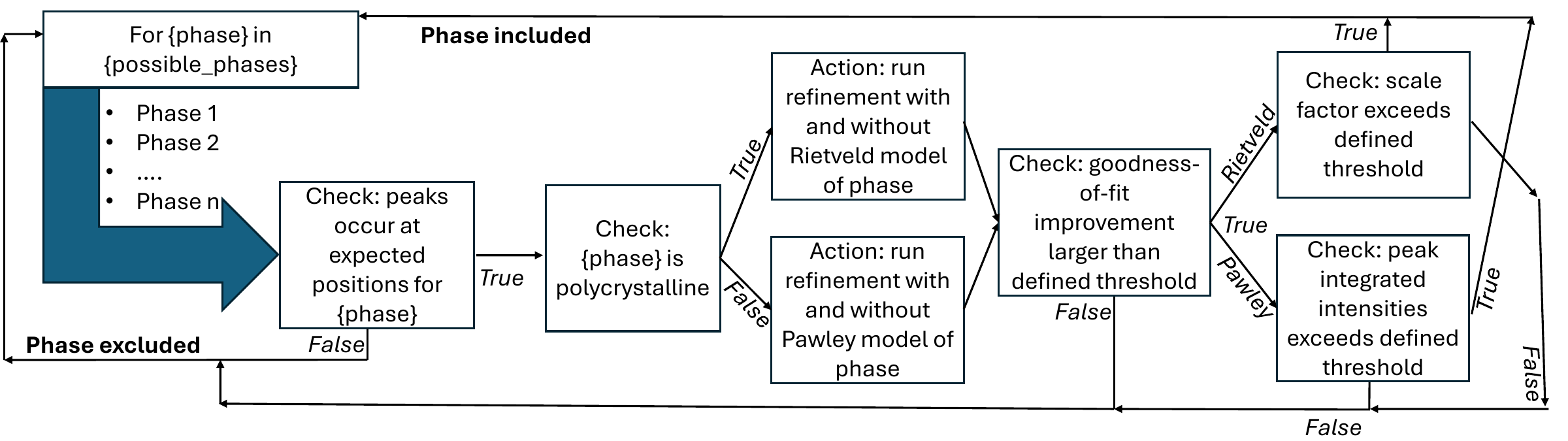}
    \caption{
        \label{protocol.pdf}
    Schematic explaining, in partial pseudo-code, the approximate phase identification algorithm used in the protocol for each diffraction pattern in a 2-dimensional grid measured over the cross-section of an SOEC. 
    }
\end{figure}

Analysis of spatially-resolved three-dimensional ($x$, $y$, and $2\theta$) diffraction data is fully automated using a sophisticated \textsc{Python 3}~\cite{van1995python} script calling \textsc{Topas 7} at the command line~\cite{Coelho2018TOPASAn}, where \textsc{Topas 7} is utilised to model the experimental data using Rietveld refinement~\cite{rietveld1969profile,rietveld2014rietveld} or, for some phases, Pawley refinement~\cite{pawley1981unit}. 
This works by taking a template \textsc{Topas} input file with a set of pre-defined phases, and running the input file multiple times with automated edits made to the file to include or exclude different phases. In our fomulation, the code also begins by performing a Rietveld refinement against a LaB$_6$ standard in order to constrain wavelength and instrumental Thompson-Cox-Hastings pseudo-Voigt~\cite{Thompson1987Rietveld} peakshape parameters but the code could easily by modified to take a different approach. 

The script does not require user-input information, such as whether the diffraction pattern in question is measured on the air or fuel electrode. 
For each diffraction pattern, the script determines which phases, amongst a series of candidate phases, are present approximately following the procedure outlined in Figure~\ref{protocol.pdf}, with some deviation for certain phases. 
This involves one or more of the following steps: 
\begin{enumerate}
    \item Check whether there are peaks matching certain characteristic peaks of a phase in the experimental diffraction pattern.
    \item Perform a Rietveld refinement with the phase in question both present and absent (but all other assigned phases present before and after refinement) and check that the improvement in goodness-of-fit (defined as the ratio of $R_\mathrm{wp}$ and $R_\mathrm{exp}$) exceeds a certain threshold value. 
    \item Perform a Rietveld refinement with the phase in question present and check that the refined scale factor of the candidate phase exceeds a certain threshold value. 
    \item Perform a Pawley refinement with the candidate phase and check that the refined peak intensities all exceed a certain threshold value. 
\end{enumerate}

For different phases, the precise assignment routine depends on morphology, x-ray scattering factor, abundance, and relationship to other phases. For most phases, the assignment algorithm consists of (1), followed by (2) and (3) if algorithm (1) affirms the possible presence of the spinel phases. 
For silver used in the current collector, the criterion for Rietveld refinement is not met because not all orientations are equally sampled, due to the large size of silver metal particles, and so for phase identification of silver, the assignment algorithm (1) and (4) is used. 

The precise details of the phase assignment routine will depend on the chemistry of the SOEC and the morphology of the different phases used in the cell. A future experiment utilising the Python code used in this study will likely need to make some modifications to the assignment algorithm due to differences in chemistry and morphology; such modifications would include changing the available phases for the algorithm to identify, and modifying the scale factor/peak intensity thresholds for phase identification. 

\subsection{Model refinement details}

\begin{table}[h!]
\centering
\caption{Summary of structural models used for phase analysis.}
\begin{tabular}{llll}
\hline
\textbf{Material} & \textbf{Space group} & \textbf{LPs (\AA)} & \textbf{Ref} \\ 
\hline
LSCF-6428 & $Pm\bar{3}m$ & $a \approx 3.89$ & \citename{Kaufman_AirElectrode_SOEC_2026} \citeyear{Kaufman_AirElectrode_SOEC_2026} \\
Co-rich spinel & $Fd\bar{3}m$ & $a \approx 8.09$ & \citename{smith1973structure} \citeyear{smith1973structure} \\

Mixed Fe/Co spinel & $Fd\bar{3}m$ & $a \approx 8.25$ & \citename{bhowmik2020high} \citeyear{bhowmik2020high} \\

Fe-rich spinel & $Fd\bar{3}m$ & $a \approx 8.40$ & \citename{fleet1981structure} \citeyear{fleet1981structure} \\

(Fe,Co)O & $Fm\bar{3}m$ & $a \approx 4.25$--$4.35$ & \citename{fjellvaag1996crystallographic} \citeyear{fjellvaag1996crystallographic} \\

8YSZ & $Fm\bar{3}m$ & $a \approx 5.146$ & \citename{batista2011structure} \citeyear{batista2011structure} \\


Ni & $Fm\bar{3}m$ & $a \approx 3.53$ & \citename{hull1921x} \citeyear{hull1921x} \\

NiO & $Fm\bar{3}m$ & $a \approx 4.196$ & \citename{barrett1964solid} \citeyear{barrett1964solid} \\

Gd$_{0.1}$Ce$_{0.9}$O$_{3-\delta}$ ($\delta\approx0.05$) & $Fm\bar{3}m$ & $a \approx 5.44$ & \citename{bandyopadhyay2021synthesis} \citeyear{bandyopadhyay2021synthesis} \\

SrO & $Fm\bar{3}m$ & $a \approx 5.142$ & \citename{primak1948x} \citeyear{primak1948x} \\
\hline
\end{tabular}
\label{tab:phase_summary}
\end{table}

Peakshapes were modelled using Thomson-Cox-Hastings pseudo-Voigt~\cite{Thompson1987Rietveld}. 
Background was fit using an order-16 Chebyshev polynomial. 
Initial applications of the protocol involved free refinement of the zero error term, however this led to unreliable trends in lattice parameters which correlated with this zero error parameter [Figure~\ref{correlation_zero-error.pdf}]; consequently zero error was fixed to 0.01$^\circ$ for all analysis presented here. 

Table~\ref{tab:phase_summary} lists the phases listed as possible phases in the model with associated symmetry and lattice parameters. 

LSCF-6428, which composes the air electrode, is typically reported to be rhombohedral~\cite{nozaki2020microstructure}. 
However, in the data collected in this study, the rhombohedral peak splitting from the cubic aristotype could not be resolved. 
Consequently use of the rhombohedral cell resulted in unstable anti-correlations between peak-width parameters and the magnitude of the rhombohedral distortion $(\sqrt{6}a_r/c_r) -1 $ (which would be zero for a cubic cell). 
The LSCF-6428 phase was therefore modelled using a cubic $Pm\bar{3}m$ unit cell ($a\approx3.89$\,\AA{}). 

Due to the large variation in (Co,Fe)$_3$O$_4$ spinel lattice parameters with composition, Fe-rich, Co-rich, and intermediate spinel was included as 3 separate possible phases. 
However, for the corresponding rocksalt (Fe,Co)O solid solution it was not necessary to use multiple phases as variation in stoichiometry has a far smaller effect on lattice parameters for (Fe,Co)O compared with the spinels [Figure~\ref{Rocksalt_LP_x.pdf}]. 

Y-stabilised zirconia (YSZ) occurs in two structures: 3YSZ (3mol\% yttria) occurs in monoclinic $P4_2/nmc$ symmetry~\cite{goulas2025formulation} and 8YSZ (8mol\% yttria) occurs in cubic $Fm\bar{3}m$ symmetry ($a\approx5.146$\,\AA{})~\cite{batista2011structure}. Although 3YSZ is used in the support layer [Figure~\ref{SOEC_diagram.pdf}], we only observe 8YSZ in the spatial range in which measurements were made in this study, and so only 8YSZ is included as a candidate phase. 

To prevent unnecessary computation, SrO, Fe/Co rocksalt, and Fe/Co spinel are only checked for at positions where LSCF-6428 is present, and Ni and NiO are only checked for if 8YSZ is present. 
This also avoids assignment of 8YSZ to SrO and vice versa, as these structures have very similar lattice parameters, x-ray scattering amplitude, and identical symmetry. 

\section{Case study}

\begin{figure}[t]
    \includegraphics[scale=0.875]{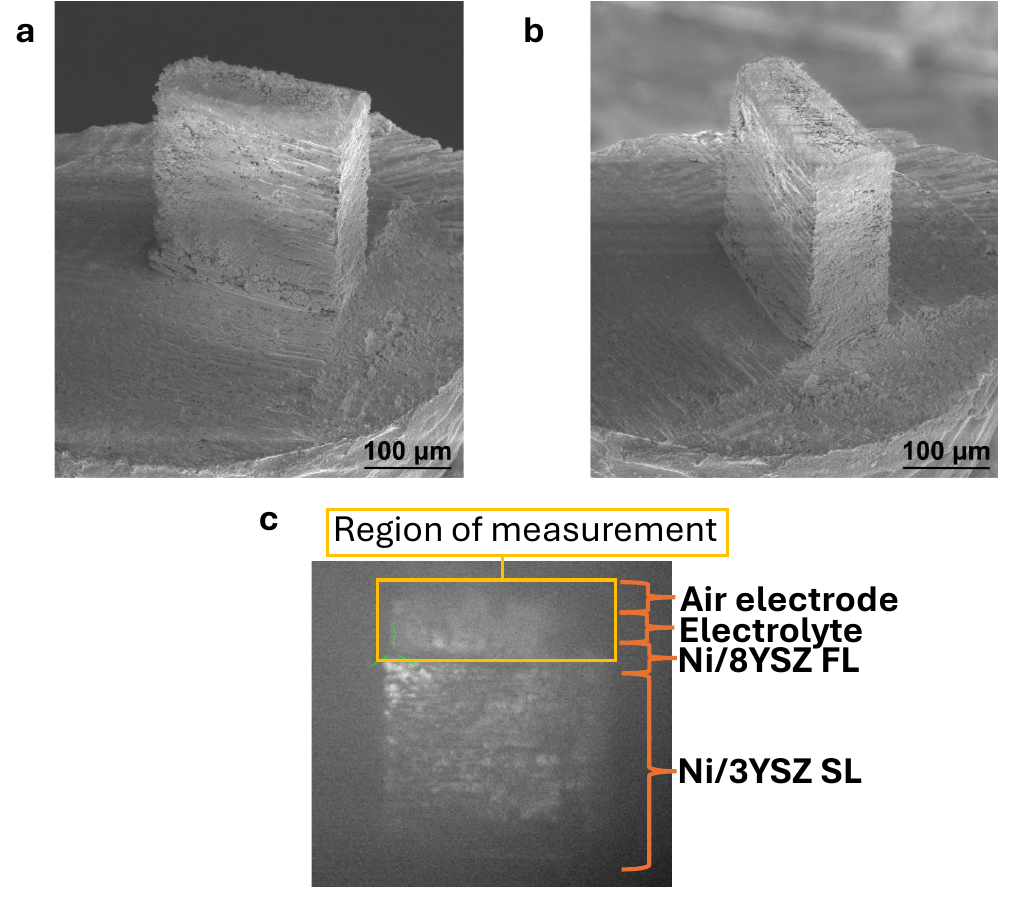}
    \caption{
        \label{tombstone_SEM_image.pdf}
    (a) and (b) SEM images of a tombstone SOEC fragment from the same cell as the fragment studied in this paper. Note that this is not the same actual fragment. 
    (c) Optical image of the tombstone SOEC fragment in the beamline, with the square indicating the region of measurement. 
    FL and SL stand for functional layer and support layer respectively.
    }
\end{figure}

SOEC button cells were cut into fragments with two different morphologies. 
One was a wide, constant-thickness shape we term a ``tombstone", the morphology of which is shown in Figure~\ref{tombstone_SEM_image.pdf} with an approximate thickness of $\sim$100\,$\mu$m. 
The other is a cylindrical shape, shown in Figure~\ref{Plasma_FIB_multiframe_BAD-sample.pdf}, around 100--200\,$\mu$m in diameter. 
These SOECs fragments were then studied at beamline 17-2~\cite{Hong2025RCI} using an anisotropic micron-focused x-ray beam. 
For the measurement on the tombstone fragments, the beam-size was 2.0\,$\mu$m in width and 1.5\,$\mu$m in height, where height is defined along the inter-electrode direction. 
This beam size is tunable and can be made narrower or broader in either spatial dimension depending on experimental necessity.  

We found during the measurements that the cylindrical morphology was not conducive to obtaining useful data, likely due to the high thickness of highly absorbing material which prevented observation of the beam from reaching the far side. We also consider that, even if the measurement had worked, the data would have been difficult to automate the analysis of, due to variable sample length resulting in very different background in resulting diffraction measurements. 
We therefore present only data from the measurements on the tombstone-shaped SOEC fragment. 

From the optical image shown in Figure~\ref{tombstone_SEM_image.pdf}(c), we can see a region to the right of the SOEC fragment where it appears there is little sample. Diffraction data in this region does show Bragg peaks associated with expected phases (i.e. LSCF-6428), however these peaks are very small relative to background [SI figure~\ref{SOEC4_10s_s01_s00070_000_C_analysis_fit_Q.pdf}]. 
To quantitatively investigate the extent to which data here is dominated by background, we calculated the fraction of the integrated diffraction data at each position which was not attributable to scattering by the sample (by integrating the Chebyshev polynomial used to model background, and integrating the total diffraction pattern). 
This is shown in Figure~\ref{SI_figure_normalised_background.pdf}. 
We see in excess of 80\% of the data dominated by background in this vicinity. 
This indicates that this region of the SOEC fragment may have been damaged before measurement, leaving very little of the cell in position in this region. 
Typically, such issues should be avoided by careful experimental preparation, but in this case we will proceed with this dataset to show how useful information can still be obtained from partially compromised data. 

\begin{figure}[t]
    \includegraphics[scale=1]{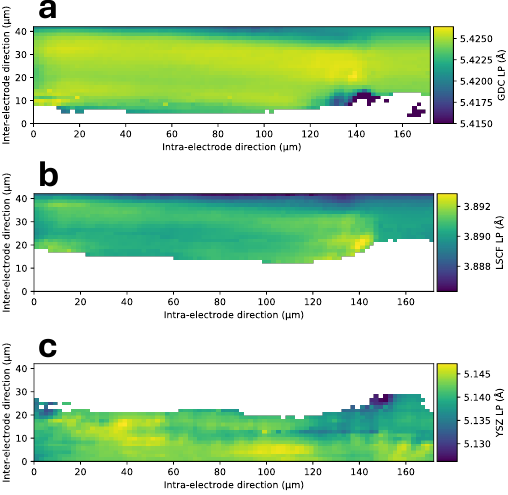}
    \caption{
        \label{SOEC4_LPs.pdf}
    Heatmap visualising the lattice parameters of (a) GDC, (b) LSCF-6428, and (c) 8YSZ as obtained by automated Rietveld refinement of diffraction data on the SOEC. 
    All three phases are modelled with cubic lattice parameters in this study. 
    White space within the heatmaps indicates pixels where the relevant phase is undetected by the algorithm, and is consistent with fuel electrode/air electrode relative positioning.
    }
\end{figure}

We have run the automated diffraction analysis over the tombstone SOEC fragment. 
Example refinements are shown in SI Figures~\ref{SOEC4_10s_s01_s00070_000_C_analysis_fit_Q.pdf}, \ref{SOEC4_10s_s02_s00004_000_C_analysis_fit_Q.pdf}, and \ref{SOEC4_10s_s28_s00074_000_C_analysis_fit_Q.pdf}. 
Figure~\ref{SOEC4_LPs.pdf} shows the spatial distribution of lattice parameters obtained by Rietveld refinement for three phases: GDC, LSCF-6428, and YSZ. 
GDC and LSCF exhibit slight decreases in lattice parameter (and hence unit cell volume) near the electrode surface. 
Unit cell volume is larger in the presence of oxygen vacancies~\cite{tyunina2021anisotropic,Kaufman_AirElectrode_SOEC_2026}, and so a reduced unit cell volume at the surface may indicate that there are fewer oxygen vacancies where the electrode makes contact with air. 
This is an intuitive result, and consistent with previous $\mu$-XRD studies on SOECs~\cite{nozaki2020microstructure}.

\begin{figure}[t]
    \includegraphics[scale=0.95]{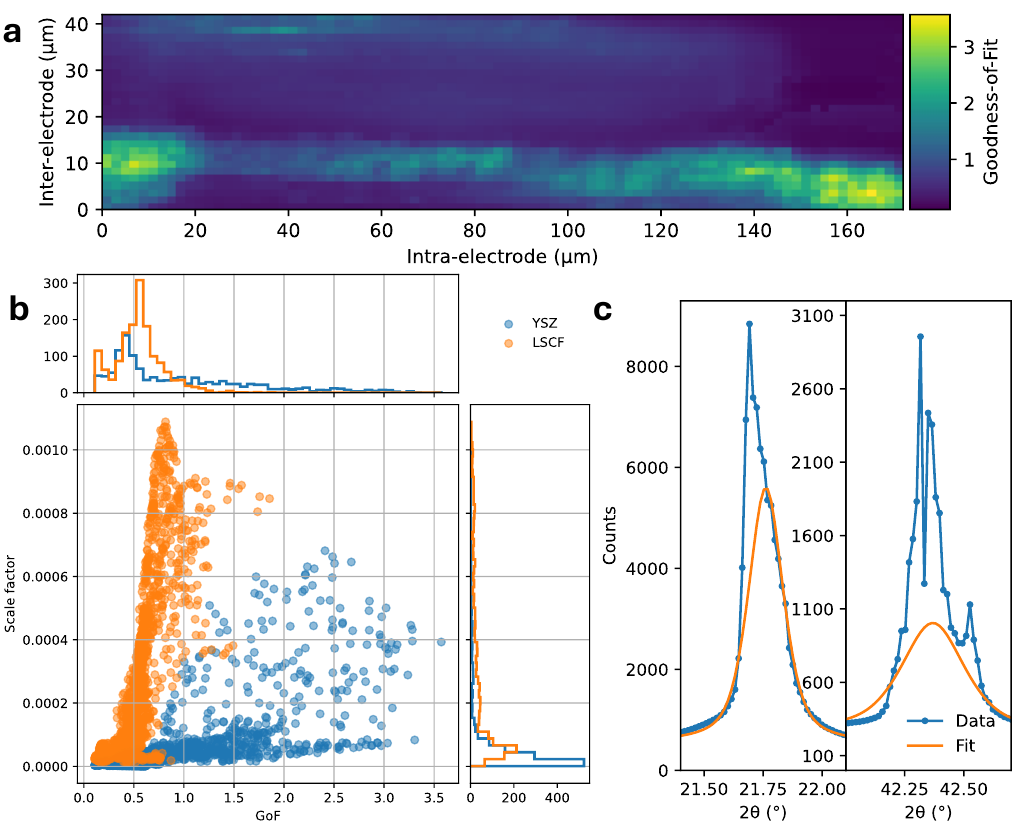}
    \caption{
        \label{YSZ-peakshape-figure.pdf}
    (a) Heatmap showing the goodness-of-fit of fits across the SOEC. 
    (b) Correlation plots of scale factor against goodness-of-fit for LSCF and 8YSZ. 
    (c) Peakshape of 8YSZ ([111] on left, [311] on right), showing the anomalous peakshape and consequent inability of standard Rietveld refinement using a Thompson-Cox-Hastings pseudo-Voigt peakshape to convincingly fit the data. 
    }
\end{figure}

Figure~\ref{YSZ-peakshape-figure.pdf}(a) shows fit quality quantified by goodness-of-fit (the ratio of $R_\mathrm{wp}$ to $R_\mathrm{exp}$) as a function of position. 
We see that fit quality is poorest in some regions of the fuel electrode. 
It is easy to perform a diagnostic on this issue using the data obtained by this automated Rietveld analysis. 
Figure~\ref{YSZ-peakshape-figure.pdf}(b) shows correlation plots of goodness-of-fit compared with 8YSZ and LSCF scale factor. 
This clearly shows that fit quality is inversely correlated with 8YSZ content, with a much stronger dependence than for LSCF, indicating that Bragg peaks associated with the 8YSZ phase are poorly modelled by the algorithm. 
While this may generally indicate an issue with the crystallographic model, in this instance we can see from Figure~\ref{YSZ-peakshape-figure.pdf}(c) that this fit issue is due to the irregular, jagged peakshapes associated with the 8YSZ phase, peakshapes which cannot easily be modelled. 
This in turn stems from a microstructure consisting of a small number of large strained grains. 
This has been seen previously in the literature, both in electron microscopy studies~\cite{sharif1998control,matsui2008grain,huang2021ebsd} and in diffraction~\cite{quach2012direct}, and stems from the high grain growth rate of 8YSZ at high temperatures. 
We have tested this in a powder-diffraction experiment detailed in SI, and observed that high-quality 8YSZ powder will coarsen with even relatively brief thermal cycling to elevated temperatures [Figure~\ref{SI_YSZ_heating_cooling.pdf}]. 
The high goodness-of-fit in the fuel electrode is therefore not a failure of the analysis protocol, but is useful information for understanding the microstructural changes occurring in an SOEC on exposure to operating temperatures. 

Additional phase-distribution heatmaps from the automated refinement are generated in Figures~\ref{SI_scale-factors-misc_air-electrode.pdf} and \ref{SI_scale-factors-misc_fuel-electrode.pdf}. 
For the air electrode [Figure~\ref{SI_scale-factors-misc_air-electrode.pdf}], the Ag contact paste/current collector is intermittently detected along the top surface. SrO is also identified in the air-electrode region, both near the outer surface facing the current collector and deeper within the electrode toward the electrolyte. 
While we believe the surface SrO assignments are likely accurate, some assignments closer to the fuel electrode may in practice reflect cubic 8YSZ which is near-identical to SrO by diffraction. 

For the fuel electrode [Figure~\ref{SI_scale-factors-misc_fuel-electrode.pdf}], Ni and NiO are detected throughout much of the electrode. 
While the presence of NiO might suggest that the reduction step was incomplete, prior \textit{operando} diffraction~\cite{Crain2026OperandoElectrolysis} and X-ray absorption spectroscopy~\cite{Zhu2026SOECDynamic} results suggest this is unlikely. 
The relative amounts of Ni and NiO are not spatially uniform: instead, the maps show distinct clusters enriched in each phase, indicating substantial local heterogeneity in the redox state of nickel across the fuel electrode. 
This observation that such reoxidised regions compose microscale agglomerates suggests they formed during lasering to prepare the fragment for $\mu$-XRD, rather than during SOEC operation, because a laser would likely induce very high local temperatures and potentially localised melting. 
To avoid this in future experiments, such lasering should be performed in inert atmosphere rather than in air, or damaged reasons should be removed by focused ion beam (FIB) . 
 
\section{Conclusion}

In this work, we have presented a fully automated analysis protocol for micron-scale, spatially resolved synchrotron X-ray diffraction data over a chemically-inhomogeneous grid. 
By combining high-flux $\mu$XRD measurements with an adaptive phase-identification strategy and automated Rietveld refinement, the protocol overcomes a major bottleneck in the analysis of large, multidimensional diffraction datasets. 
Crucially, the approach does not require prior knowledge of local device architecture or manual intervention during analysis, enabling robust and scalable treatment of thousands of diffraction patterns acquired across complex material cross-sections.

Application of the protocol to operated solid oxide electrolyser cells demonstrates its ability to robustly analyse spatially heterogeneous diffraction data and to extract physically meaningful trends across multiple functional layers. 
With no user input regarding the relative positions of the air electrode and fuel electrode, the protocol successfully identifies their positions and hence the relevant phases. 
The automated workflow captures systematic variations in crystallographic parameters, in particular the tendency for decreased unit cell volume (and hence decreased O vacancy content) in proximity to air, consistent with previous results from \citename{nozaki2020microstructure} \citeyear{nozaki2020microstructure}. 
These results validate the protocol as a practical and informative tool for \textit{post mortem} analysis of complex, multi-phase electrochemical devices. 

More broadly, the analysis framework presented here is readily extensible to other electrochemical and energy-conversion devices, including solid oxide fuel cells, all-solid-state batteries, and photovoltaic materials, where spatially heterogeneous phase evolution is common. While the protocol represents a significant step toward fully autonomous analysis of spatially resolved diffraction data, there remains scope for further development, for example through the incorporation of machine learning or artificial intelligence approaches for phase recognition, anomaly detection, or adaptive model selection, building directly on the structured and physically grounded outputs of the present method. With modest modification of candidate phase lists and assignment criteria, the protocol can be adapted to different chemistries and morphologies. By reducing the barrier to comprehensive, spatially resolved diffraction analysis, this work is expected to facilitate wider adoption of $\mu$XRD as a routine diagnostic tool and to accelerate the development and durability optimization of these devices.

Finally, we believe that while the approach detailed here is an important stepping stone, there remains significant scope to improve upon our approach. A bottleneck with our approach is the need to arbitrarily define threshold scale factors and goodness-of-fit improvements for phase assignments, and further to define the phases that might possibly occur in a dataset before beginning the analysis. With the advent of machine learning and large language models, such user-defined steps may in future be eliminated, further streamlining the process of analysis of complex multi-dimensional diffraction datasets. 

\section{Experimental}

\subsection{Cell fabrication and testing}

Details of SOEC button cell fabrication and operation have been reported previously~\cite{zhu2025voltage,slomski2026structure,Zhu2026SOECDynamic}. 
Briefly, the cell architecture consists of a 500\,$\mu$m thick Ni-YSZ support layer, a 20\,$\mu$m thick fine Ni-YSZ active fuel electrode layer, an $\sim$7 $\mu$m dense YSZ electrolyte, a 5\,$\mu$m Gd$_{0.1}$Ce$_{0.9}$O$_{1.95}$ (GDC) barrier layer, a 20 $\mu$m porous 50/50 GDC/LSCF (La$_{0.6}$Sr$_{0.4}$Co$_{0.2}$Fe$_{0.8}$O$_{3-\delta}$) air electrode and a porous LSCF contact of varying thickness. 
Layers are screen printed and fired in successive steps to make the cell, then contacted and reduced on the testing rig. 
It was cooled and removed after reduction and prior to any long-term testing to serve as a beginning-of-life baseline. 


\subsection{Specimen Preparation and Imaging}

Specimens were isolated from the SOECs using an Oxford Lasers A Series micromachining tool (532\,nm Laser 10 -- 500\,ns pulses, max average power 1.5\,W). First, a 1\,mm diameter disk was removed from the cell by scribing a circle into the Ni-YSZ support layer at 0.375\,W over 500 passes at 2\,mm/s until the disk was cut loose from the sample bulk. The $\sim$1\,mm disk was then adhered to a 1\,mm diameter stainless steel dowel using UV-cure resin. Initial thinning of the specimen was performed operating the laser as a lathe at 0.75\,W to reduce the diameter to $\sim$0.5\,mm over a length of $\sim$0.35\,mm to encompass a portion of each electrode and the internal layers.

Rectangular specimens were then shaped to final dimensions by scoring two 0.2\,mm lines spaced $\sim$0.2\,mm apart along the axial dimension (perpendicular to the electrolyte/electrode interfaces). Then, lines were scored perpendicular to the initial cut from the edge of the sample to the initial axial cut to remove excess specimen. This process was repeated after rotating the sample 90$^\circ$, with the only difference that the initial scoring lines were placed $\sim$0.1\,mm apart, yielding the rectangular prism specimen with dimensions of $\sim$ 0.2\,mm$\times$0.1\,mm$\times$0.2\,mm. SEM images of the rectangular specimen were acquired using a Nova Nano 630 SEM (Thermo Fisher Scientific, Waltham MA, USA) operated at 5\,kV with a 1.8\,nA beam current. Secondary electron (SE) imaging mode was used with an Everhart-Thornley detector (ETD).

Columnar specimens were shaped to near-final dimensions by performing a second thinning process targeting a final diameter of $\sim$0.3\,mm using a lathing process at 0.25\,W. Due to the presence of a laser damaged surface region following laser lathing, a final thinning step was performed on a Thermo Fisher Scientific Helios 5 Laser Hydra Plasma-FIB-SEM using a Xe ion beam operated at 30\,kV with 1\,$\mu$A beam current. SEM images of the polished columns were acquired on the same instrument operated at 5\,kV with 1.6\,nA beam current. Full-specimen images were acquired in secondary electron (SE) mode with the ETD detector and higher magnification images of the cell layers were acquired in SE mode with the TLD through-lens detector.

\subsection{Synchrotron x-ray diffraction}

SSRL beamline 17-2 is a high brightness in-vacuum undulator insertion device, featuring a multilayer monochromator for enhanced flux. 
In the $\mu$XRD configuration, mirrors with acceptance of 9 milliradians in the vertical plane and 4.5\,milliradians in the horizontal plane are used to focus the beam to a spot size on the order of 100s of microns in vertical and horizontal dimensions upstream of the sample. The beam then diverges and is refocused by secondary Kirkpatrick-Baez mirrors to a small beam size. 
This step size for the spatially-resolved measurements were $2.0 \times 1.0$\,$\mu$m for the measurements on the tombstone-shaped SOEC fragments, and $2.4 \times 3.4$\,$\mu$m for the measurements with the cylindrical SOEC fragments; these step sizes were roughly informed by the beam profile [Figure~\ref{SI_beam-profile.pdf}] to minimize pixel overlap while avoiding undersampling. 

An optical camera is focused to the X-ray focal point with a tolerance of approximately $\pm5$\,$\mu$m. The leading surface of the SOEC cylindrical posts was initially aligned to the optical axis parallel to the X-ray beam, followed by mapping the horizontal and vertical bounds of the sample using translations of the sample in the direct beam. Mapping arrays were achieved by scanning the sample horizontally across the width of the sample with step sizes equal to the beam width. Following the completion of a line scan, a vertical translation equal to the vertical beam profile was performed and the horizontal line scan repeated. Mapping was performed until a translation of at least 80\,$\mu$m relative to the air-LSCF interface.

Two-dimensional X-ray diffraction images were measured with an Eiger 4M hybrid photon counting detector positioned $\sim$235\,mm from the sample position, as calibrated by a NIST 660c LaB$_6$ standard. 2D images were normalized by incident beam intensity and sample transmission as determined by a photodiode beam stop positioned behind the sample and subsequently integrated into 1D patterns using the Nika macro~\cite{ilavsky2012nika} within Igor Pro.

\begin{acknowledgements}
L. A. V. N-C thanks John S. O. Evans (Durham University) for useful discussions. 
All graphs prepared using \textsc{MatPlotLib}~\cite{Hunter:2007} implemented in Python~\cite{van1995python}. 
\end{acknowledgements}

\begin{funding}
This work was authored in part by the National Laboratory of the Rockies for the U.S. Department of Energy (DOE) under Contract No. DE-AC36-08GO28308.  
Use of the Stanford Synchrotron Radiation Lightsource, SLAC National Accelerator Laboratory, is supported by the U.S. DOE, Office of Science, Office of Basic Energy Sciences under Contract No. DE-AC02-76SF00515. 
The views expressed in the article do not necessarily represent the views of the DOE or the U.S. Government. 

M.G.P.'s work is supported in part by the U.S. Department of Energy, Office of Science, Office of Workforce Development for Teachers and Scientists (WDTS) under the Science Undergraduate Laboratory Internships (SULI) Program.
\end{funding}

\ConflictsOfInterest{None to declare.}

\DataAvailability{The code prepared for this project, and data used in this analysis, are available for use in a public repository on \textsc{GitHub} at https://github.com/lnaglecocco/muXRD-2026~\cite{naglecocco_muxrd2026}. Reuse of the code, with or without modification, is welcomed, with credit given by citing this paper.}

\bibliography{iucr} 

\clearpage
\appendix
\section*{Supplementary Information}
        \renewcommand{\thesection}{\arabic{section}}
        \renewcommand{\theequation}{S\arabic{equation}}
	\renewcommand{\thetable}{S\arabic{table}}
	\renewcommand{\thefigure}{S\arabic{figure}}
	\setcounter{equation}{0}
	\setcounter{section}{0}
	\setcounter{figure}{0}
	\setcounter{table}{0}

\section{Electron microscopy}

\begin{figure}[hbtp]
    \includegraphics[width=\linewidth]{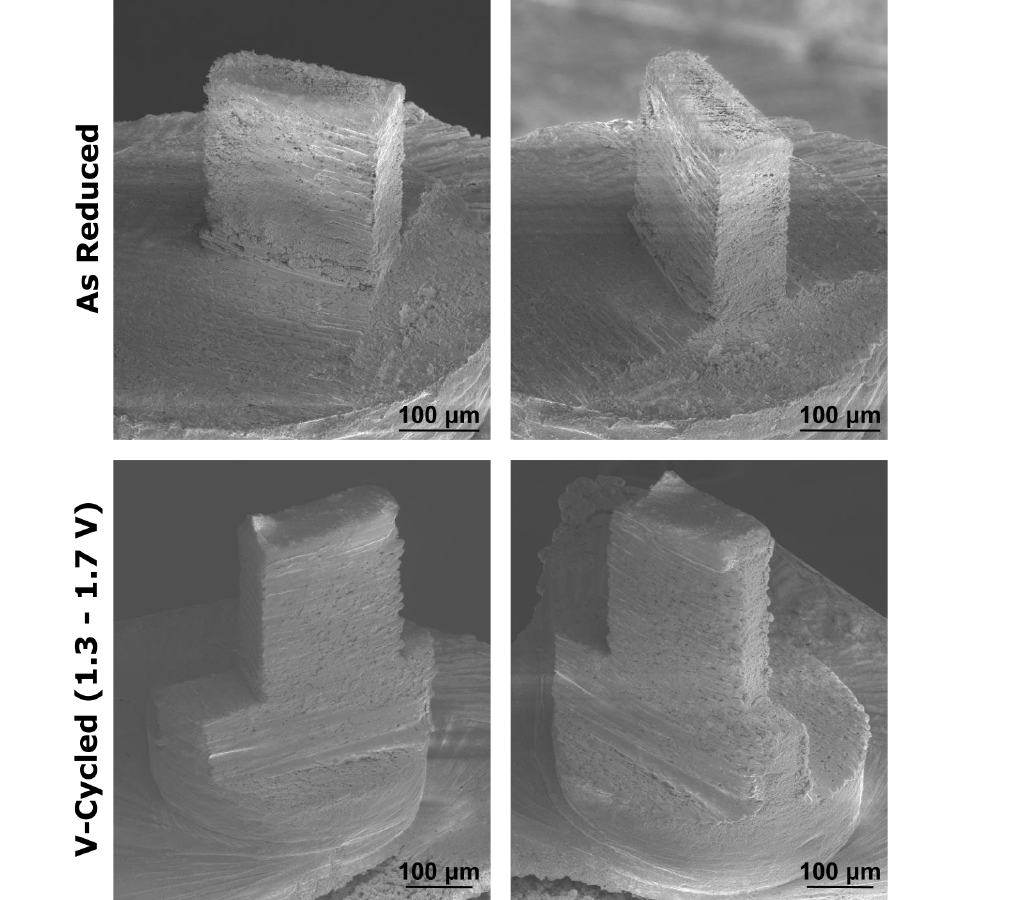}
    \caption{
        \label{SEM_tombstone_SI}
       Image of tombstone-shaped SOEC fragments. 
       Top: an SOEC fragment from the cell investigated in this study. 
       Bottom: an SOEC fragment from a voltage-cycled cell which may be the subject of a future work. 
    }
\end{figure}

\begin{figure}[p]
    \includegraphics[scale=0.1]{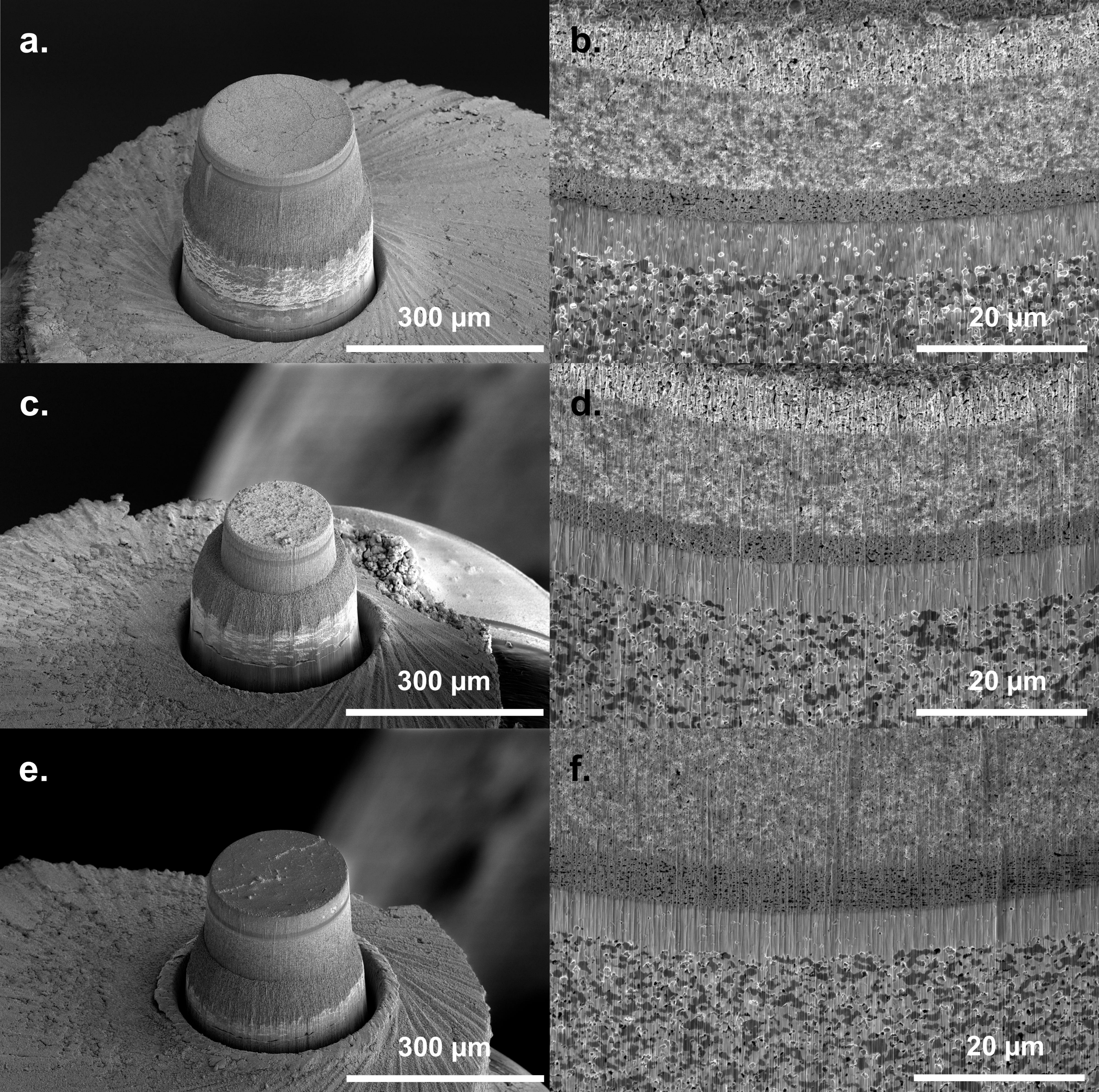}
    \caption{
        \label{Plasma_FIB_multiframe_BAD-sample.pdf}
       Image of cylinder-like SOEC fragments which proved unsuitable for spatially-resolved cross-sectional $\mu$-XRD due to the large resulting path length through highly-absorbing material, and the variable sample width. 
    }
\end{figure}

In this section we present electron microscopy images of the solid oxide electrolyser cell fragments studied in this work. 
Figure~\ref{Plasma_FIB_multiframe_BAD-sample.pdf} shows the cylindrical SOEC fragments we measured, which did not yield useful diffraction due to the high absorption and inconsistent path length. 
Figure~\ref{SEM_tombstone_SI} shows SOEC fragments with the ``tombstone" morphology studied in this work.

\clearpage\section{Optical images of SOEC fragments}

\begin{figure}[hbtp]
    \includegraphics{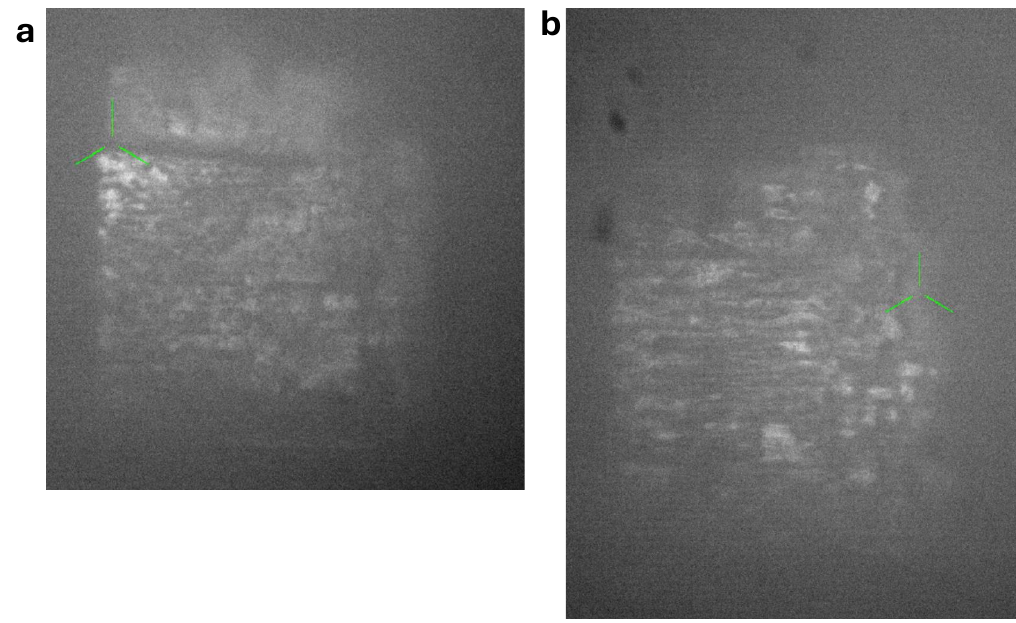}
    \caption{
        \label{SI_tombstones_optical.pdf}
       Optical images of two example SOEC fragments in tombstone configuration, as observed within beamline 17-2. 
       (a) is the sample studied in this work, and (b) is a different SOEC fragment. 
    }
\end{figure}

In this section, we present in Figure~\ref{SI_tombstones_optical.pdf} the optical images of the tombstone SOEC fragments studied in this work.

\clearpage\section{Beam profile}

\begin{figure}[hbtp]
    \includegraphics[scale=0.95]{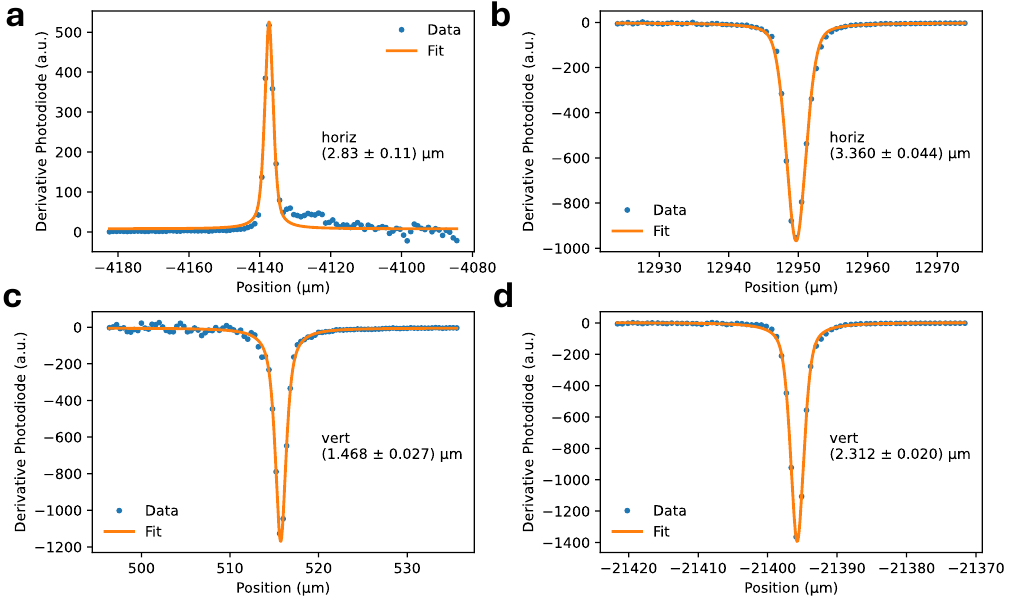}
    \caption{
        \label{SI_beam-profile.pdf}
        The focused beam width, as determined by fitting a pseudo-Voigt peakshape, for the (a,b) horizontal and (b,d) vertical beam directions, for the (a,b) tombstone SOEC fragments and (c,d) cylindrical SOEC fragments.
    }
\end{figure}

The focused beam profile was characterised by scanning a sharp edge through the beam and fitting the derivative of the resulting photodiode signal with a pseudo-Voigt function using \textsc{SciPy}~\cite{2020SciPy-NMeth}. 
The resulting fits are shown in Figure~\ref{SI_beam-profile.pdf}. For the tombstone SOEC fragments, the horizontal beam width was $2.83 \pm 0.11\,\mu$m (Figure~\ref{SI_beam-profile.pdf}a) and the vertical beam height was $1.468 \pm 0.027\,\mu$m (Figure~\ref{SI_beam-profile.pdf}c). 
For the cylindrical fragments, the corresponding dimensions were $3.360 \pm 0.044\,\mu$m horizontally (Figure~\ref{SI_beam-profile.pdf}b) and $2.312 \pm 0.020\,\mu$m vertically (Figure~\ref{SI_beam-profile.pdf}d). 
The slightly increased beam dimensions for the cylindrical specimens reflect differences in optical configuration and focusing conditions between measurement sessions, a consequence of the difficulty in achieving such high focus. 
These experimentally determined beam sizes were used to roughly inform the step size in spatial mapping, ensuring minimal undersampling while avoiding excessive oversampling of the illuminated volume.

\clearpage\section{Results of refining zero error}

\begin{figure}[p]
    \includegraphics[width=0.8\linewidth]{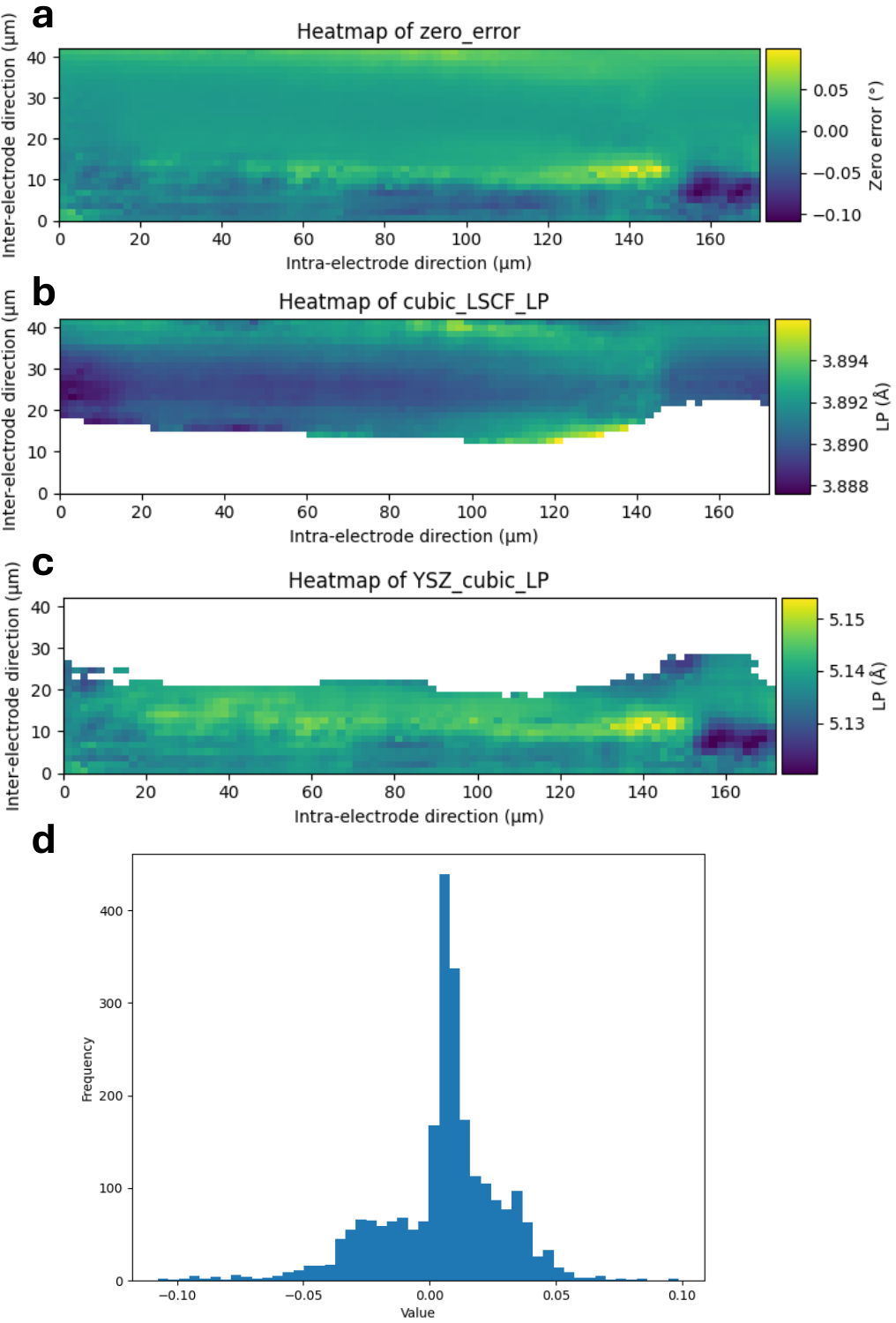}
    \centering
    \caption{
        \label{correlation_zero-error.pdf}
       A figure showing the correlation between lattice parameters and zero error, if zero error is allowed to refine. 
       To avoid this outcome, zero error should not be allowed to refine freely during analysis. 
       Note that the rest of the analysis in this manuscript uses fixed zero error. 
       (a) A heatmap over the SOEC showing the variation in zero error. 
       (b) A heatmap over the SOEC showing variation in YSZ cubic lattice parameter, which clearly correlates with refined zero error. 
       (c) A heatmap over the SOEC showing variation in LSCF-6428 cubic lattice parameter, which clearly correlates with refined zero error. 
       (d) A histogram of refined zero errors over all pixels in this analysis. 
    }
\end{figure}

In the results presented in this work, we keep zero error fixed across all positions. 
While this may represent an issue if surface texture results in variation in the beam path length to the sample as a function of position, we ultimately found that refining zero error resulted in problematic correlations between lattice parameters and zero error, as shown in Figure~\ref{correlation_zero-error.pdf}.

\clearpage\section{Dependence of lattice parameter on Fe:Co content for metal oxides}

\begin{figure}[hp]
    \includegraphics{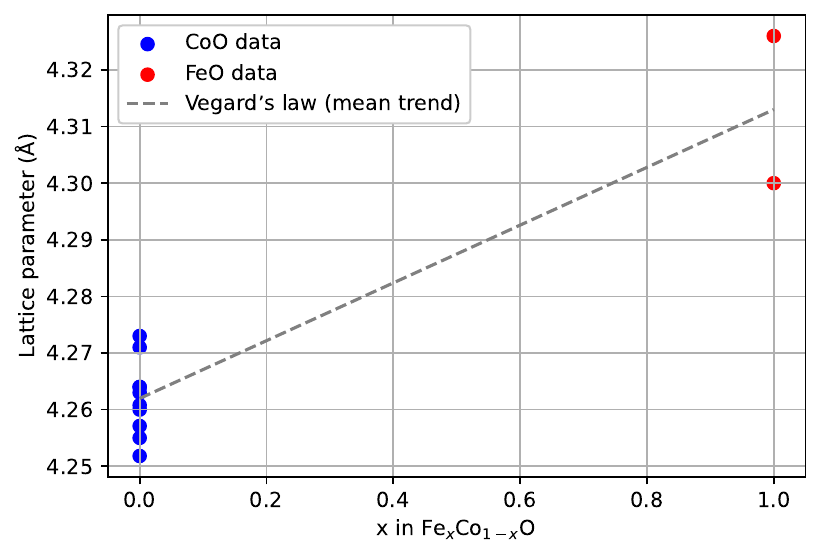}
    \caption{
        \label{Rocksalt_LP_x.pdf}
       Dependence of lattice parameter on composition for the FeO-CoO rocksalt solid solution assuming Vegard's law behaviour, using various end-member lattice parameters reported on the Inorganic Crystal Structure Database (collection codes: CoO 128532, 9865, 53058, 191776, 245319, 245320, 245321, 245322, 245323, 245324; FeO 82233, 633038). 
    }
\end{figure}

\begin{figure}[hp]
    \includegraphics[width=\linewidth]{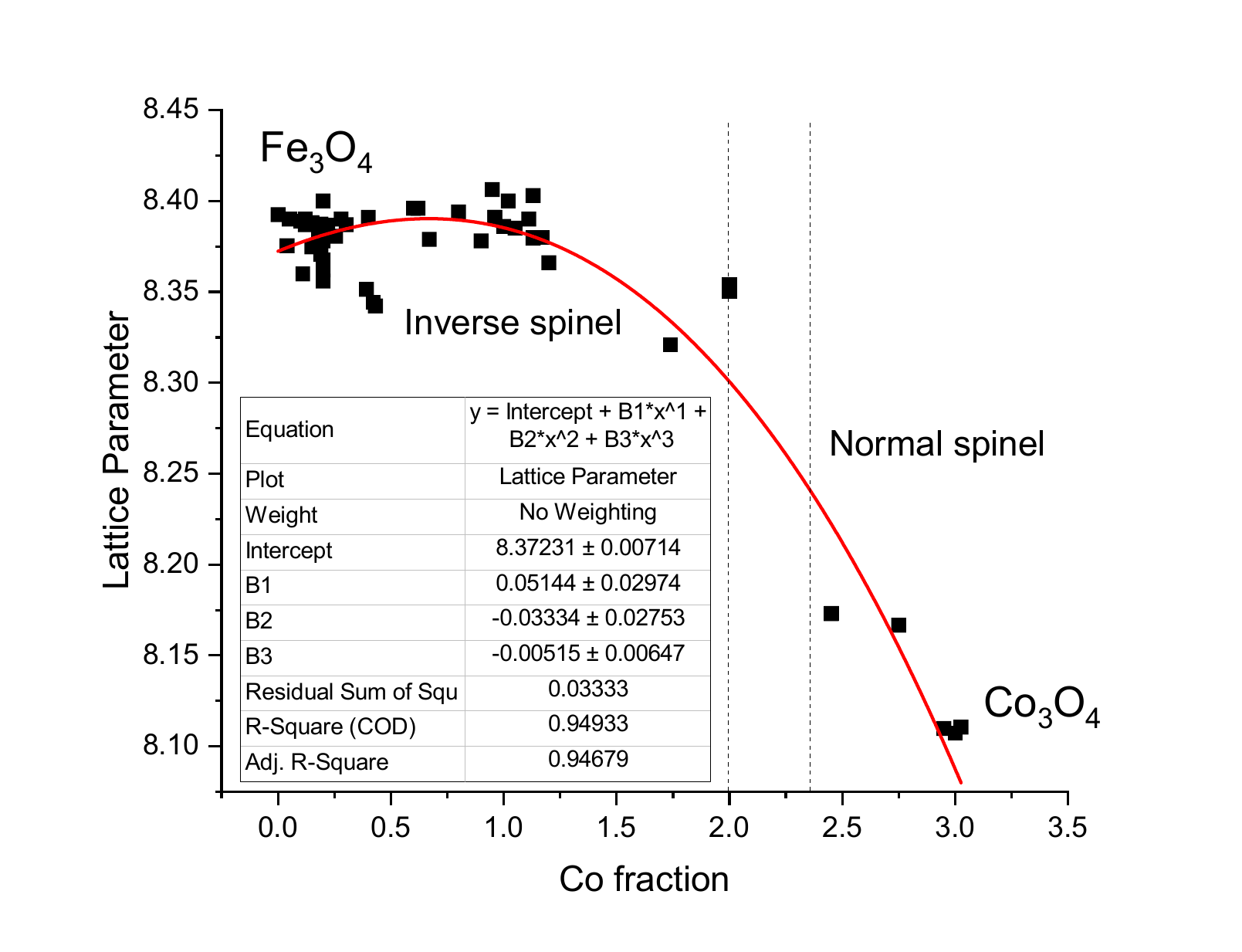}
    \caption{
        \label{spinel_LP_x.pdf}
       Dependence of lattice parameter on composition for the Fe$_3$O$_4$-Co$_3$O$_4$ spinel solid solution, using various sets of lattice parameters reported on the Inorganic Crystal Structure Database. 
       In the table shows the results of a fit using an order-3 polynomial, demonstrating that Vegard's law is not followed here.
    }
\end{figure}

As discussed in the main text, spinels $M_3$O$_4$ and rocksalts $M$O (where $M$ is a combination of Fe and Co) exhibit different dependence of lattice parameter on stoichiometry. 
Figure~\ref{Rocksalt_LP_x.pdf} shows the dependence for rocksalt and Figure~\ref{spinel_LP_x.pdf} shows the dependence for spinels, using various data from the Inorganic Crystal Structure Database. 

\clearpage\section{Example diffraction analysis}

\begin{figure}[hbtp]
    \includegraphics[width=\linewidth]{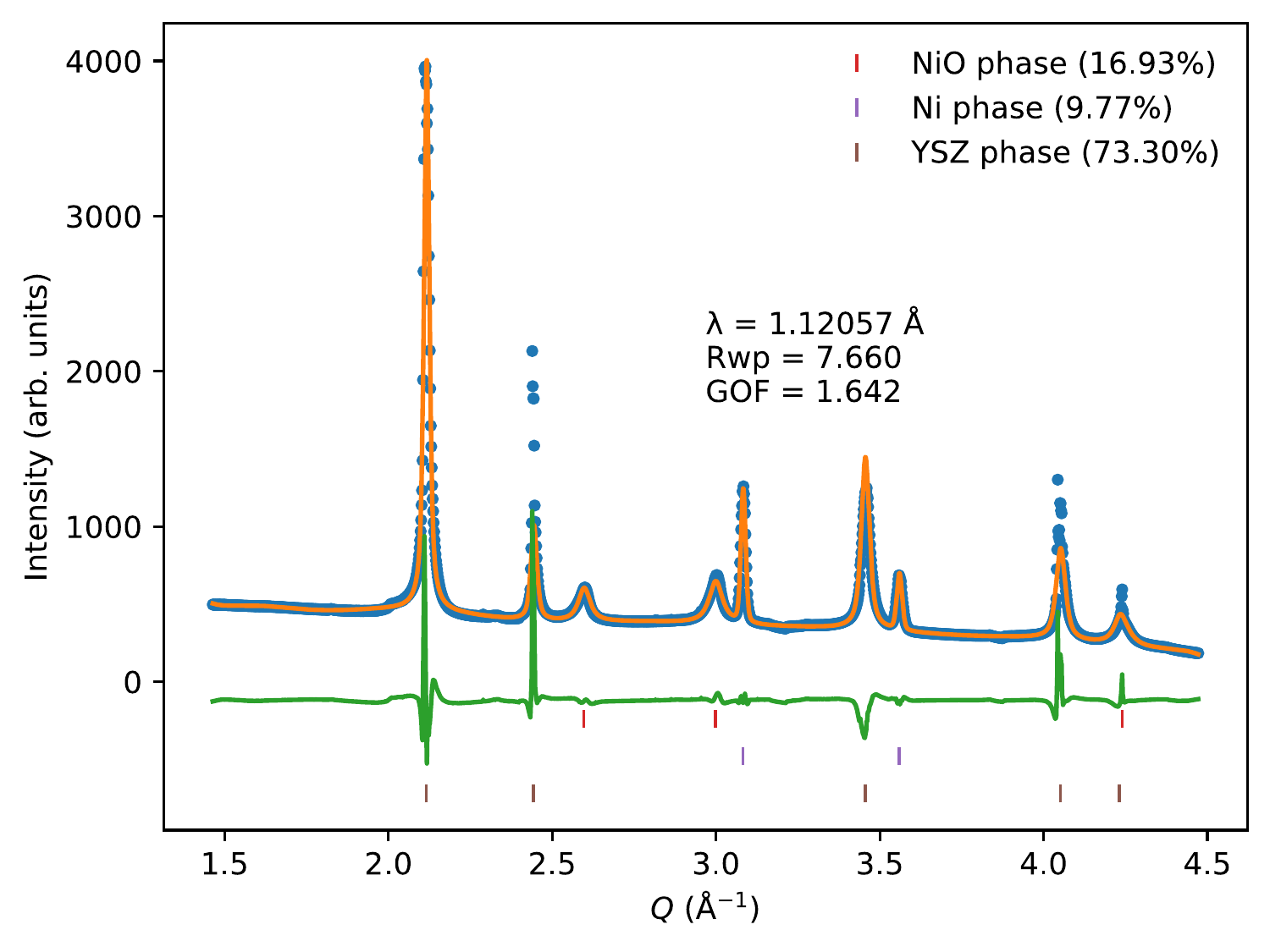}
    \caption{
        \label{SOEC4_10s_s28_s00074_000_C_analysis_fit_Q.pdf}
       An example diffraction pattern at the fuel electrode of the SOEC fragment, in the 10th row and 75th 
       column of the measurement grid, with associated Rietveld fit using the phases identified by the protocol. 
       Ni and NiO coexist with 8YSZ, with 8YSZ giving rise to spikes in the difference pattern due to the phase morphology as discussed in the main text of the manuscript.
    }
\end{figure}

\begin{figure}[hbtp]
    \includegraphics[width=\linewidth]{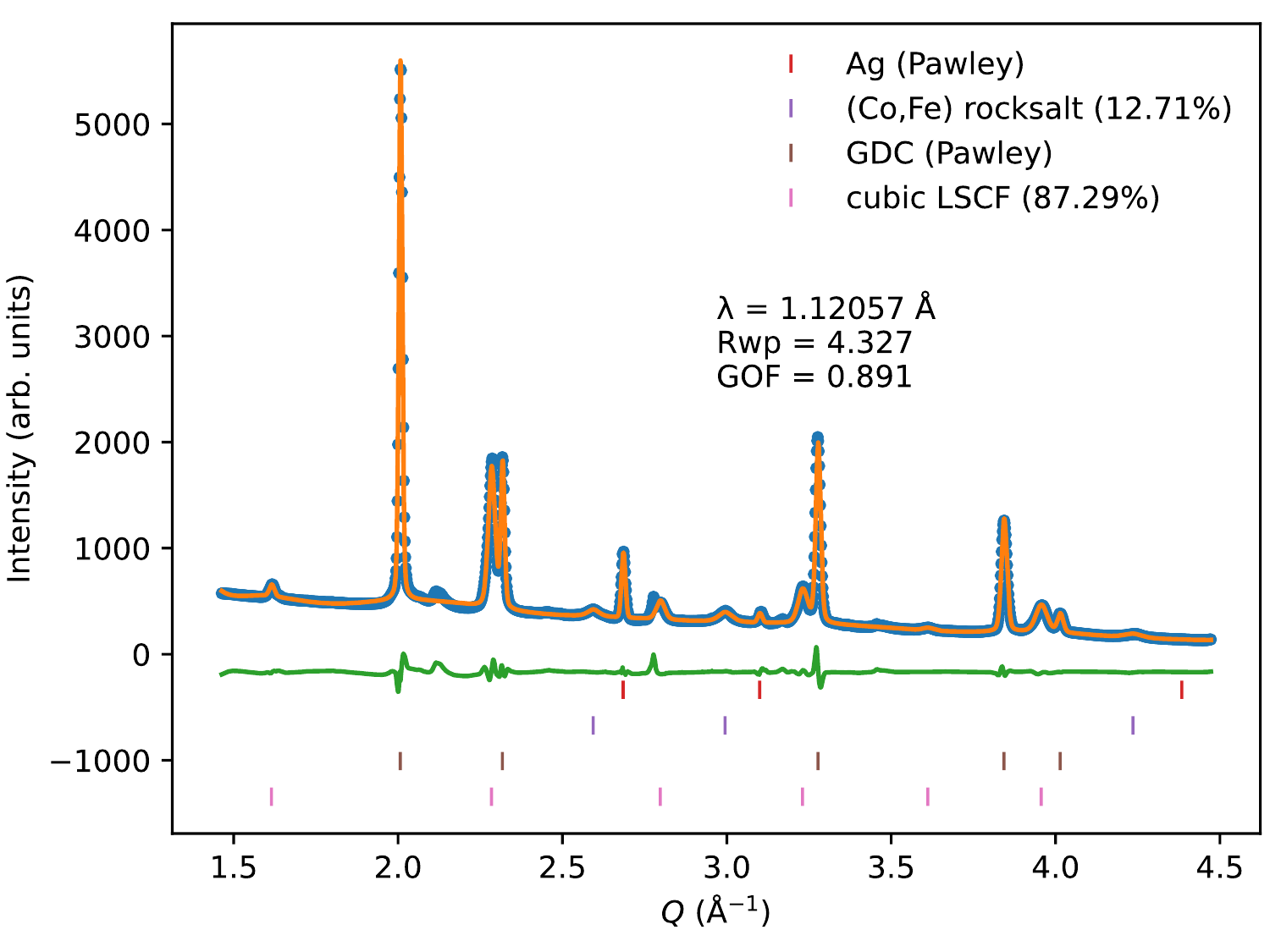}
    \caption{
        \label{SOEC4_10s_s02_s00004_000_C_analysis_fit_Q.pdf}
       An example diffraction pattern at the air electrode of the SOEC fragment, in the 2nd row and 5th 
       column of the measurement grid, with associated Rietveld and Pawley fit using the phases identified by the protocol. 
    }
\end{figure}

\begin{figure}[hbtp]
    \includegraphics[width=\linewidth]{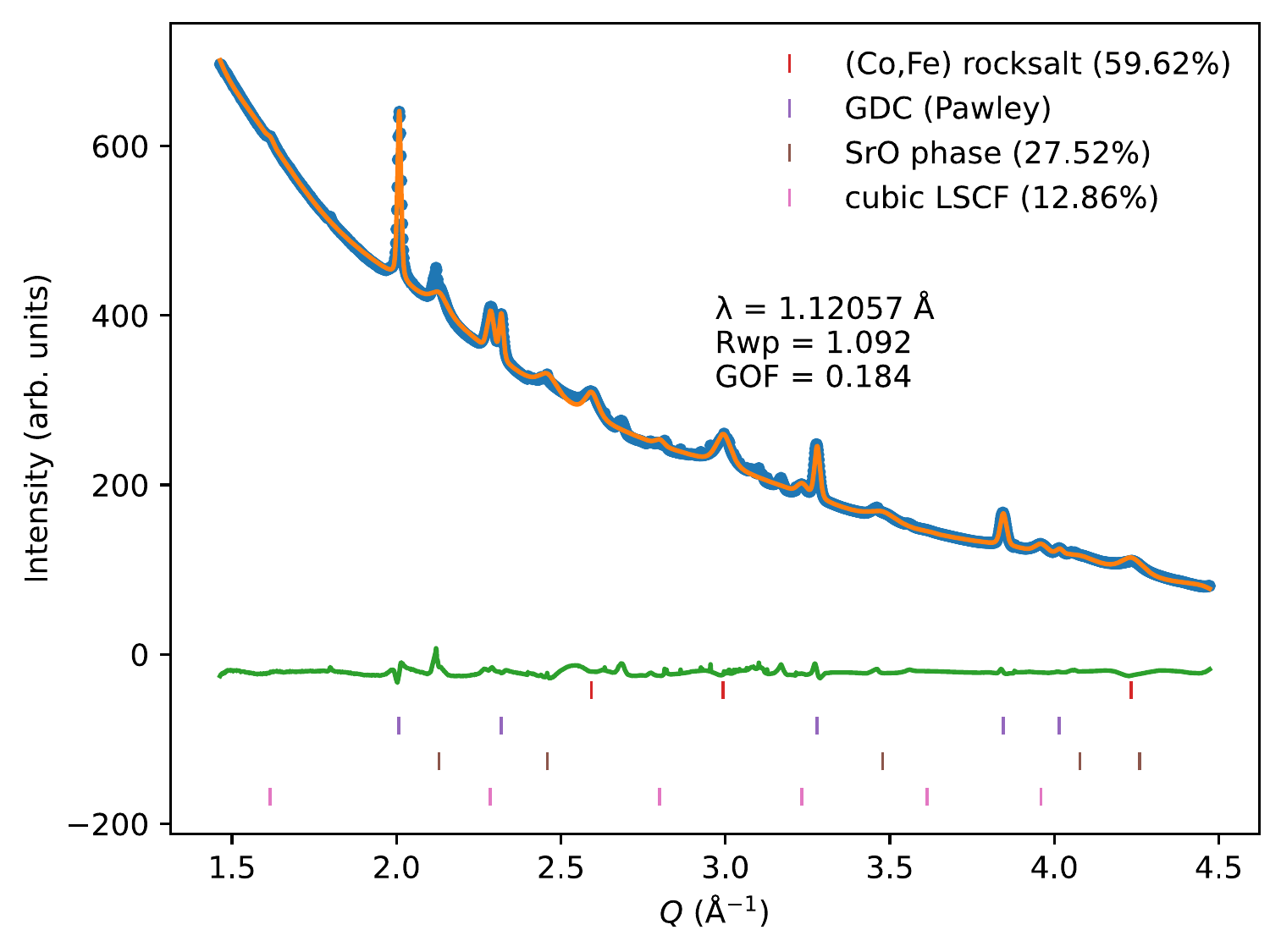}
    \caption{
        \label{SOEC4_10s_s01_s00070_000_C_analysis_fit_Q.pdf}
       An example diffraction pattern at the air electrode of the SOEC fragment, in the 1st row and 71st 
       column of the measurement grid, with associated Rietveld and Pawley fit using the phases identified by the protocol. 
       This occurs in a region where we believe the SOEC fragment partially crumbled, as suggested by the optical image in Figure~\ref{SI_tombstones_optical.pdf}, which is why the data is dominated by background effects. 
       This background domination is reflected by Figure~\ref{SI_figure_normalised_background.pdf} which is a heatmap showing the fraction of diffraction data which is background. 
    }
\end{figure}

In this section, we show some example final refinements performed by the protocol described in this work. 
Figures~\ref{SOEC4_10s_s28_s00074_000_C_analysis_fit_Q.pdf}--\ref{SOEC4_10s_s01_s00070_000_C_analysis_fit_Q.pdf} present representative diffraction patterns and associated refinements from different regions of the SOEC fragment. 
The first example, Figure~\ref{SOEC4_10s_s28_s00074_000_C_analysis_fit_Q.pdf}, from the fuel electrode, shows a Rietveld fit in which Ni and NiO coexist with 8YSZ, with the 8YSZ phase producing characteristic spikes in the difference curve due to its morphology. 
The second and third examples are taken from the air electrode and illustrate combined Rietveld and Pawley fits using the protocol-identified phases. While the second air-electrode region exhibits a well-resolved pattern, the third [Figure~\ref{SOEC4_10s_s01_s00070_000_C_analysis_fit_Q.pdf}] corresponds to a partially crumbled region of the fragment, leading to strong background contributions that dominate the diffraction signal, consistent with the optical observations and the background fraction heatmap, Figure~\ref{SI_figure_normalised_background.pdf}, discussed elsewhere.

\clearpage\section{Heatmaps of SOEC from refinement algorithm}

In this section, we show some additional outputs of our automated analysis protocol. 

\begin{figure}[hbtp]
    \includegraphics[scale=0.8]{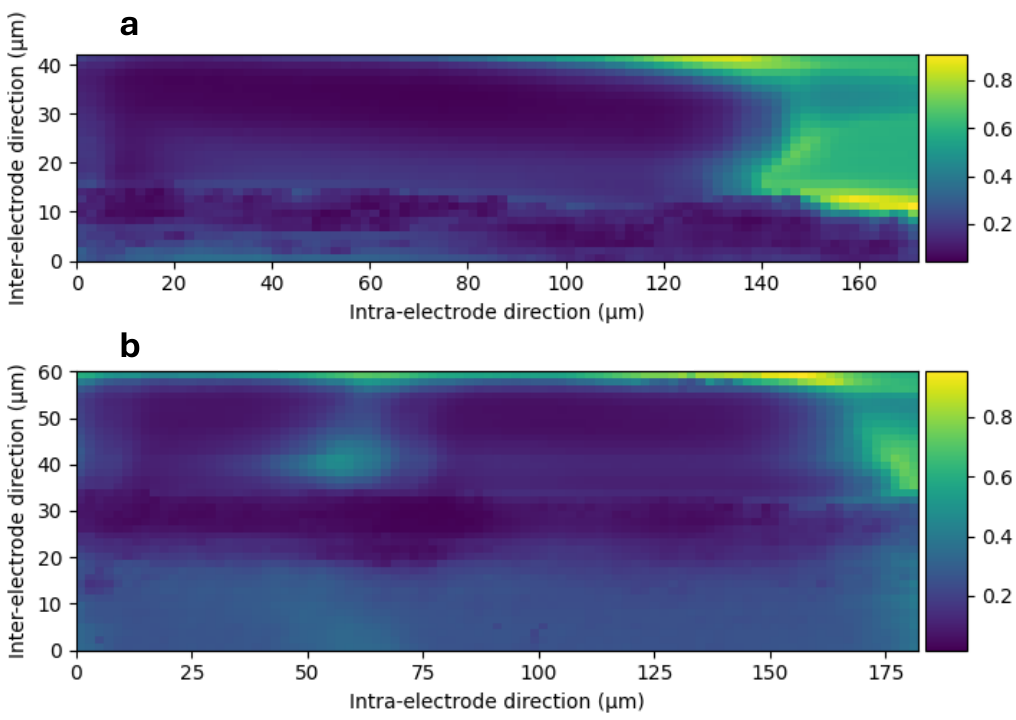}
    \caption{
        \label{SI_figure_normalised_background.pdf}
       The ratio between the integrated background (which is modelled in the Rietveld algorithm as a Chebyshev polynomial) and the total integrated data for (a) the SOEC fragment studied in this work and (b) another SOEC fragment measured in the same experiment. 
       Regions where the ratio is very high indicate low data quality, likely due to low sample quality. 
    }
\end{figure}

Figure~\ref{SI_figure_normalised_background.pdf} shows the fraction of data which is due to background; a high ratio indicates poor data quality. 
This figure is consistent with the optical image shown earlier [Figure~\ref{SI_tombstones_optical.pdf}] which indicates that part of the SOEC fragment may have crumbled. 

\begin{figure}[hbtp]
    \includegraphics[width=\linewidth]{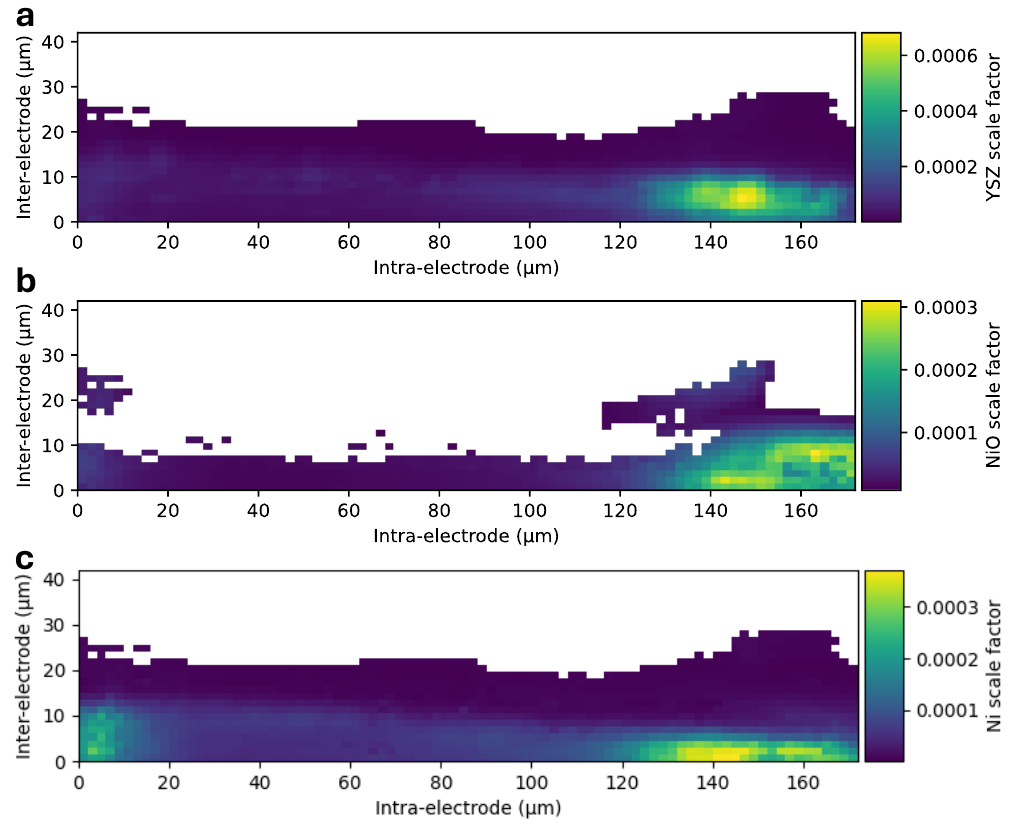}
    \caption{
        \label{SI_scale-factors-misc_fuel-electrode.pdf}
       The Rietveld scale factors obtained for the fuel electrode phases (a) 8YSZ, (b) NiO, and (c) Ni metal, according to the automated protocol in this work, for a tombstone SOEC fragment in which air electrode is at the top and fuel electrode is at the bottom. 
    }
\end{figure}

\begin{figure}[hbtp]
    \includegraphics[width=\linewidth]{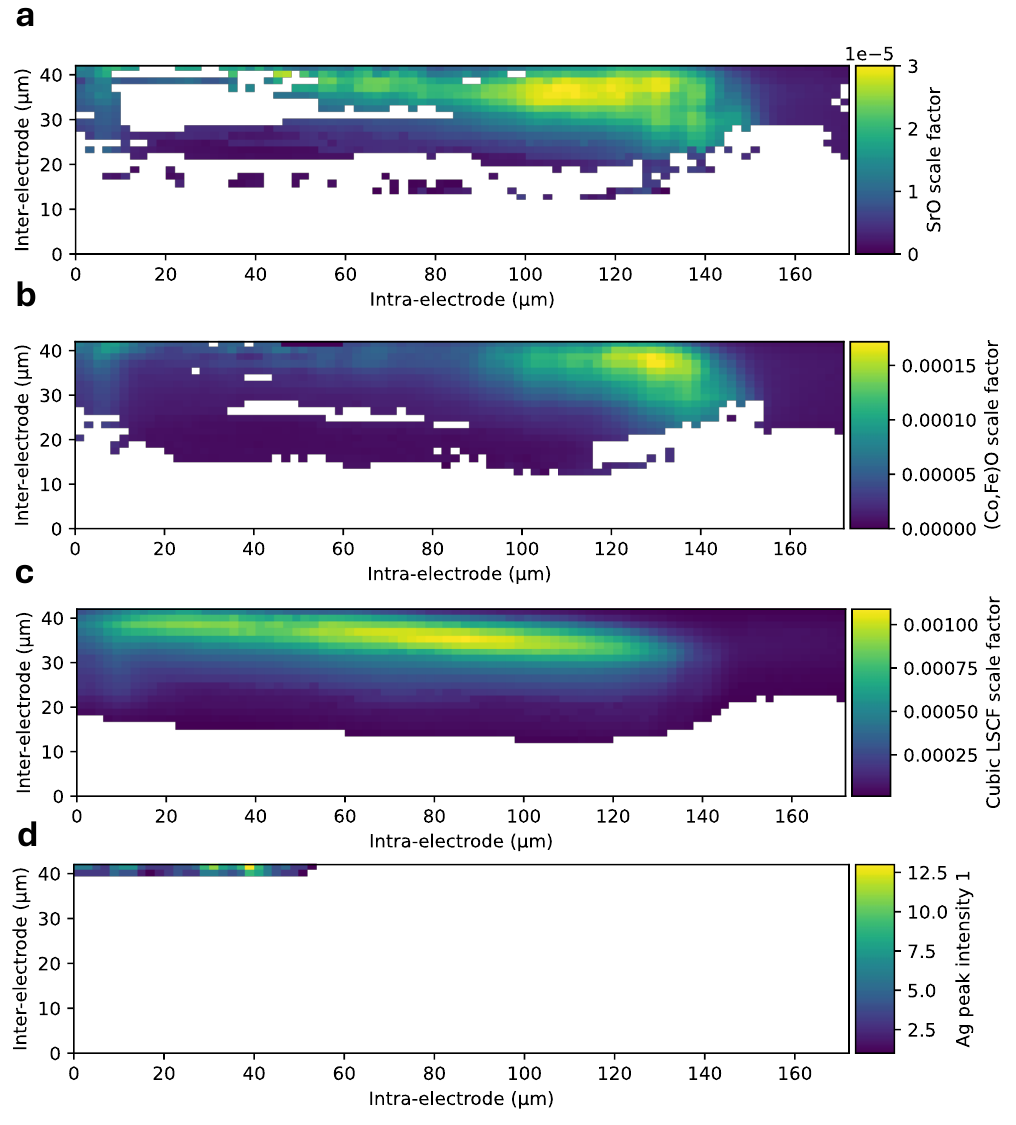}
    \caption{
        \label{SI_scale-factors-misc_air-electrode.pdf}
       The Rietveld scale factors obtained for the air electrode phases (a) SrO, (b) Co$_x$Fe$_{1-x}$O rocksalt, and (c) LSCF-6428. 
       (d) shows the intensity of the first Bragg peak associated with the Ag contact paste, visible for some part of the surface of the sample, as determined by Pawley refinement for that phase. 
       These results determined by the automated protocol in this work, for a tombstone SOEC fragment in which air electrode is at the top and fuel electrode is at the bottom. 
    }
\end{figure}

Figure~\ref{SI_scale-factors-misc_fuel-electrode.pdf} shows the distributions of scale factor for the phases that occur at the fuel electrode. 
Figure~\ref{SI_scale-factors-misc_air-electrode.pdf} shows the distributions of scale factor or (for silver, which is treated by Pawley approach) integrated peak intensity for the phases that occur at the air electrode.

\clearpage\section{Powder diffraction experiment on 8YSZ}

To understand the grain coarsening observed in 8YSZ observed in the micron-focused spatially-resolved diffraction experiment in this work, we performed an \textit{in situ} heating experiment on beamline 2-1 at SSRL.

\subsection{Methods for powder diffraction experiment}

Cerium (IV) oxide with 20mol\% Gd dopant (GDC) was purchased from Sigma Aldrich, with advertised particle size $<100$\,nm. 
ZrO$_2$ stabilised with 8mol\% Y$_2$O$_3$ (YSZ), with proclaimed particle size D$_{50}$ in the range 3--5\,$\mu$m, was purchased from Fuel Cell Materials. 
These two materials were mixed in a 50:50 ratio by mass.

An aliquot of this GDC-YSZ mixture was placed in a 1.0\,mm outer diameter, 0.58\,mm inner diameter quartz capillary and measured on BL2-1~\cite{stone2023remote} at SSRL using an x-ray beam of energy 17\,keV. 
Heating was supplied using resistive coils mounted to a sample cell described elsewhere~\cite{hoffman2018situ}, with temperature recorded using a K-type thermocouple. 
The sample was heated to 750$^\circ$C, held for 2\,hrs, and cooled to room-temperature, measured continuously throughout this process.

\subsection{Results}

\begin{figure}[hbtp]
    \includegraphics[width=\linewidth]{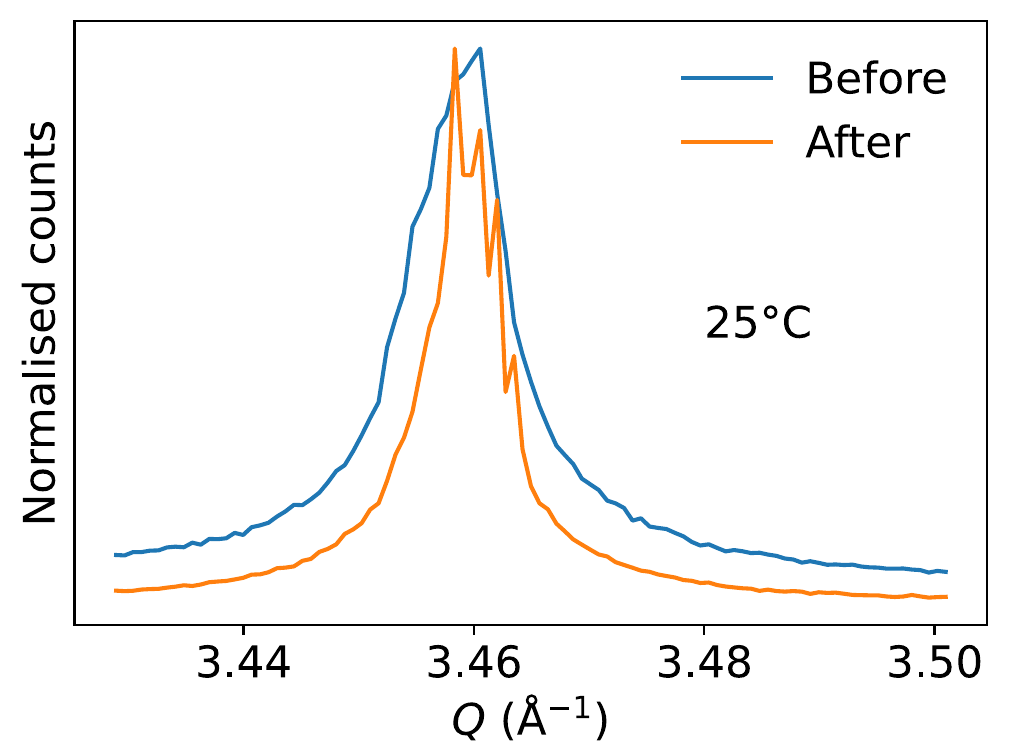}
    \caption{
        \label{SI_YSZ_heating_cooling.pdf}
       8YSZ [220] peak at 25$^\circ$C before and after the heating/cooling cycle, in a powder diffraction experiment at beamline 2-1 where GDC and 8YSZ were mixed in a 50:50 mass ratio. 
    }
\end{figure}

The 8YSZ [220] peak at room temperature is shown in Figure~\ref{SI_YSZ_heating_cooling.pdf} before and after a thermal cycle to 750$^\circ$C. 
There is clear evidence of inhomogeneous grain growth, both in terms of narrowing of the peak, and reduction in peakshape quality due to inhomogeneous grain size. 
This explains the failure of the analysis protocol to accurately model the 8YSZ peakshape, as shown in Figure~\ref{YSZ-peakshape-figure.pdf}.

\end{document}